\documentclass[
aps,
pre,
floatfix,
twocolumn
]{revtex4-1}
\usepackage{graphicx}
\usepackage{amsmath,amssymb}
\usepackage{float}
\usepackage{color}
\usepackage{graphics}
\usepackage{epsfig}

\makeatletter
\@ifundefined{textcolor}{}
{
 \definecolor{BLACK}{gray}{0}
 \definecolor{WHITE}{gray}{1}
 \definecolor{RED}{rgb}{1,0,0}
 \definecolor{GREEN}{rgb}{0,1,0}
 \definecolor{BLUE}{rgb}{0,0,1}
 \definecolor{CYAN}{cmyk}{1,0,0,0}
 \definecolor{MAGENTA}{cmyk}{0,1,0,0}
 \definecolor{YELLOW}{cmyk}{0,0,1,0}
}

\makeatother

\begin{document}

\title{Symmetry breaking in competing single-well linear-nonlinear potentials%
}
\author{Dmitry A. Zezyulin$^1$}
\author{Mikhail E. Lebedev$^{2,3}$}
\author{Georgy L. Alfimov$^{2,4}$}
\author{Boris A. Malomed$^{5,1}$}

\affiliation{$^1$ITMO University, St.~Petersburg 197101, Russia}

\affiliation{$^2$Institute of Mathematics with Computer Center, Ufa Scientific
	Center, Russian Academy of Sciences, Chernyshevskii str., 112, Ufa 450008,
	Russia}

\affiliation{$^3$All-Russian Institute for Scientific and Technical
	Information, Russian Academy of Sciences, 20 Usievich str, Moscow, 125190,
	Russia}

\affiliation{$^4$Moscow Institute of Electronic Engineering,
	Zelenograd, Moscow, 124498, Russia}

\affiliation{$^5$Department of Physical Electronics, School of Electrical
	Engineering, Faculty of Engineering, and Center for Light-Matter Interaction,
	Tel Aviv University, Tel Aviv 69978, Israel}

\date{\today }

\begin{abstract}
The combination of linear and nonlinear potentials, both shaped as a single
well, enables competition between the confinement and expulsion induced by
the former and latter potentials, respectively. We demonstrate that this
setting leads to spontaneous symmetry breaking (SSB) of the ground state in
the respective generalized nonlinear Schr\"{o}dinger (Gross-Pitaevskii)
equation, through a spontaneous off-center shift of the trapped mode. Two
different SSB bifurcation scenarios are possible, depending on the shape of
the nonlinearity-modulation profile, which determines the nonlinear
potential. If the profile is bounded (remaining finite at $|x|\rightarrow
\infty $), at a critical value of the integral norm the spatially symmetric
state loses its stability, giving rise to a pair of mutually symmetric
stable asymmetric ones via a direct pitchfork bifurcation. On the other
hand, if the nonlinear potential is unbounded, two unstable asymmetric modes
merge into the symmetric metastable one and destabilize it, via an inverted
pitchfork bifurcation. Parallel to a systematic numerical investigation,
basic results are obtained in an analytical form. The settings can be
realized in Bose-Einstein condensates and nonlinear optical waveguides.
\end{abstract}

\maketitle


\section{Introduction}
\label{sec:intro}
External potentials, which steer the propagation and
trapping of electromagnetic fields in photonics and matter waves in
Bose-Einstein condensates (BECs), feature spatial symmetry in many
physically important settings, a commonly known example being double-well
potentials, which were explored in diverse physical settings \cite{DWP},
often in a combination with nonlinearity \cite{Walls}-\cite{Zegadlo} (see
also book \cite{JYang}). It is well known too that the symmetry of the
ground state in models including the self-focusing nonlinearity follows the
symmetry of the underlying potential structure, which is always true in
quantum mechanics \cite{LL} (and, more generally, in any linear theory),
only as long as the nonlinearity remains weak enough. A generic effect,
which sets in at a critical strength of the nonlinearity, is \textit{%
spontaneous symmetry breaking} (SSB). In particular, in the framework of the
nonlinear Schr\"{o}dinger equation (NLSE), which is a generic model for many
settings in optics and BEC, SSB was first considered in Ref. \cite{Davies}
(in terms of a discretized version of the NLSE, the SSB\ concept was
introduced in another early work \cite{self}). Generally, above the SSB
point, the ground state of the one-dimensional (1D) NLSE model amounts to a
soliton which is located at one of local minima of the underlying double- or
multi-well potential structure. Because the particular minimum is chosen
spontaneously among two or several available ones, the respective ground
state features double or multiple degeneracy, which is possible in nonlinear
systems, being forbidden in the linear Schr\"{o}dinger equation \cite{LL}.

In addition to the usual linear (in particular, double-well) potentials,
NLSE-based models may include nonlinear \textit{pseudopotentials} \cite%
{pseudo}, which are represented by a spatially-dependent coefficient in
front of the nonlinear term {(we stress that the nonlinear interaction 
remains local, while its strength may become a function of coordinates)}. 
Pseudopotentials can be created in optics, by implanting
nonlinearity-inducing dopants into the host medium \cite{Kip}, as well as in
BEC, making use of a locally applied Feshbach resonance (FR), controlled by
tightly focused laser beams. Several examples of spatial and spatiotemporal
nonlinear pseudopotentials for BEC\ have been created in the experiment by
means of the latter technique \cite{FR1}-\cite{FR3}. Another possibility is
the creation of an effective pseudopotential ``painted" by a
rapidly moving laser beam \cite{Painting}. Various soliton modes supported
by pseudopotentials have been studied in detail (thus far, in the
theoretical form) \cite{Barcelona}. In particular, double-well
pseudopotentials based on the self-attractive nonlinearity give rise to
specific SSB effects \cite{Thaw}-\cite{Dror}.

In this work we aim to demonstrate that the SSB is possible not only in
double-well (pseudo)potentials, but also in a combination of \emph{competing}
linear and nonlinear \emph{single-well} potentials, assuming that the former
one is confining, while the nonlinear potential is expulsive. {In
terms of optics, this may be considered as a combination of a linear
waveguide and nonlinear antiwaveguide. Previously, the opposite situation
was considered, viz., competition of linear antiwaveguiding and nonlinear
self-focusing, which produced not SSB effects, but transient regimes for
quasi-stable propagation of spatial solitons \cite{anti}.} {To the
best of our knowledge, the SSB in the absence of a double-well structure was
not demonstrated previously. Here, we find that,} depending on the shape of
the nonlinear potential, the SSB proceeds according to one of the two
possible scenarios: in one case, the original symmetric ground state loses
its stability through a direct bifurcation of a pitchfork type \cite{bif},
which gives rise to a pair of stable asymmetric states; in the other case,
the pitchfork bifurcation is inverted, leading to merger of two asymmetric
states into a metastable symmetric one, leading to its destabilization.

The rest of the paper is organized as follows. In Sec.~\ref{sec:model}, we describe the physical model. In Sec.~\ref{sec:numerical}, we present the main numerical results concerning the SSB in the competing linear-nonlinear potentials. Section~\ref{sec:analytical} offers the analytical treatment that supports the numerical results. Finally, Sec.~\ref{sec:conclusion} concludes the paper and provides an outlook on   future research.

\section{The model}
\label{sec:model}
We consider the 1D NLSE, written here as the
Gross-Pitaevskii equation (GPE)\ which governs the evolution of the
macroscopic BEC\ wave function $\Psi (x,t)$ \cite{GPE}:
\begin{equation}
i\Psi _{t}=-\Psi _{xx}+\frac{1}{2}\omega ^{2}x^{2}\Psi -P(x)\Psi |\Psi |^{2},
\label{1D_GP}
\end{equation}%
with time $t$ and coordinate $x$ scaled so that $\hbar =1$ and normalized
atomic mass is $m=1/2$, whereas $\omega ^{2}$ is the strength of the linear
harmonic-oscillator (HO)\ trapping potential, and $P(x)>0$ determines the
nonlinear pseudopotential, induced by the self-attractive nonlinearity,
which we adopt in the form of a simple single-well structure:
\begin{equation}
P(x)=1+A\tanh ^{2}x,  \label{eq:bounded}
\end{equation}%
with $A>0$. {In BEC, this pseudopotential profile can be easily
created \ by juxtaposing a local FR, induced by a focused laser beam, with
the uniform self-attraction controlled by the uniform illumination, which
can be readily implemented, e.g., in the condensate of }$^{174}${Yb
atoms \cite{FR2}. As for the linear HO potential, it is a standard
ingredient of any experimental setting dealing with BEC. In optics, the
local-nonlinearity modulation profile defined by Eq. (\ref{eq:bounded}) can
be made by uniformly doping the periphery of the waveguide with a resonant
nonlinearity-enhancing material \cite{Kip}, while leaving the core area
undoped. }

As shown below, it is important that function $P(x)$ is bounded, i.e., its
maximum value, $P(x=\pm \infty )=1+A$, is finite. Several other profiles $%
P(x)$ with shapes similar to one defined by Eq. (\ref{eq:bounded}) have also
been considered, to check that the results reported below do not essentially
depend on the specific choice of the bounded pseudopotential. {This
conclusion implies that the results are structurally stable, once they are
not affected by a variation of the particular shape of the pseudopotential
(obviously, in the actual experiment the shape can be designed with a finite
accuracy).}

The Hamiltonian corresponding to Eq. (\ref{1D_GP}) is%
\begin{equation}
H=\int_{-\infty }^{+\infty }\left[ \left\vert \Psi _{x}\right\vert ^{2}+%
\frac{1}{2}\omega ^{2}x^{2}\left\vert \Psi \right\vert ^{2}-\frac{1}{2}%
P(x)\left\vert \Psi \right\vert ^{4}\right] dx.  \label{H}
\end{equation}%
It follows from here that the structure defined by Eq. (\ref{eq:bounded})
represents a repulsive nonlinear-potential barrier, as the strength of the
local self-attraction has a minimum at $x=0$. Accordingly, the equilibrium
position of a soliton in the present system is determined by the competition
between the trapping linear HO potential and expulsive pseudopotential
corresponding to Eq. (\ref{eq:bounded}). The subsequent analysis reveals the
main finding of this work: the competition of the linear and nonlinear
spatially symmetric potentials creates an \textit{asymmetric} ground state,
past the SSB point, {in the absence of any double-well potential.} We
also demonstrate that switching from the bounded pseudopotential (\ref%
{eq:bounded}) to an unbounded one leads to a significantly different SSB
scenario, with neither symmetric state nor asymmetric one being stable past
the bifurcation point.

We look for stationary states with real chemical potential $\mu $ as $\Psi
(x,t)=e^{-i\mu t}u(x)$, where $u(x)$ is a localized real stationary wave
function which satisfies equation
\begin{equation}
u_{xx}+\mu u-\frac{1}{2}\omega ^{2}x^{2}u+P(x)u^{3}=0,  \label{eq:stat}
\end{equation}%
and can be characterized by its norm (proportional to the number of atoms in
the corresponding BEC), $N=\int_{-\infty }^{+\infty }u^{2}(x)dx$. Symmetric
(even) and antisymmetric (odd) solutions are defined, respectively, by $%
u(x)=u(-x)$ and $u(x)=-u(-x)$, whereas for asymmetric modes one has $%
|u(x)|\neq |u(-x)|$.



\section{Numerical results}
\label{sec:numerical}
Symmetric and symmetry-broken states can be
found numerically, applying the standard iterative Newton's method, with a
properly chosen initial guess, to Eq. (\ref{eq:stat}). Since in this work we
are interested in fundamental nodeless solutions, we used the initial guess
in the form of Gaussian profiles $u_{0}(x)=a\exp \left(
-(x-x_{0})^{2}\right) $, where $a$ and $x_{0}$ are the trial parameters
(obviously, $x_{0}=0$ should be chosen if the target solution is symmetric,
and $x_{0}\neq 0$ otherwise). If the Newton's method converges to a
stationary solution for some $\mu $, a family of nonlinear modes can be
found by means of the continuation in $\mu $. To characterize the SSB
bifurcation, we introduce the center-of-mass coordinate,%
\begin{equation}
X_{c}=N^{-1}{\int_{-\infty }^{+\infty }xu^{2}(x)dx},  \label{Xc}
\end{equation}%
which is zero for symmetric and antisymmetric solutions, and nonzero for
asymmetric ones.

\begin{figure}[tbp]
\includegraphics[width=0.99\columnwidth]{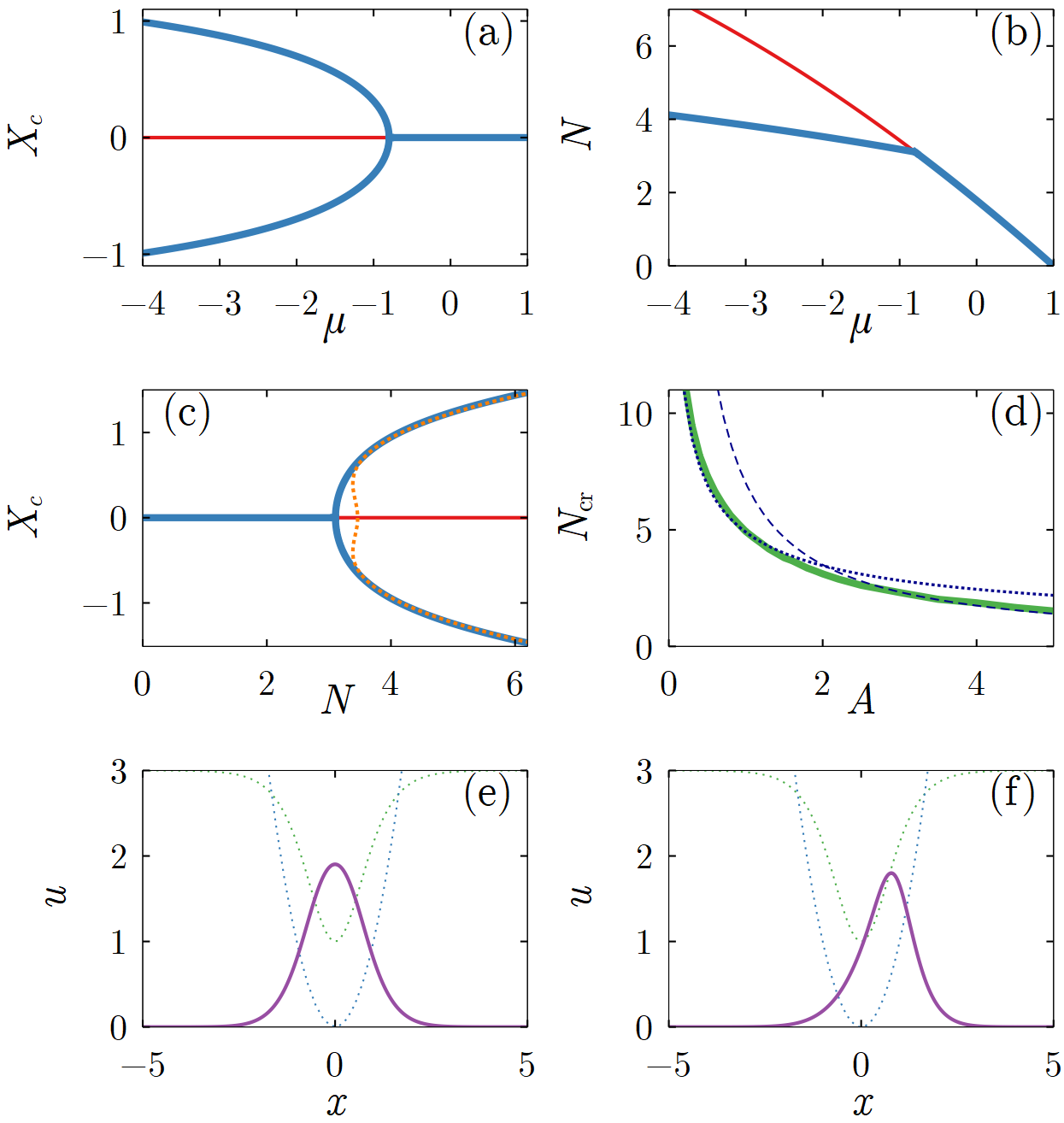}
\caption{Panels (a,b,c) show three renditions of the SSB bifurcation for
bounded pseudopotential (\protect\ref{eq:bounded}) with $A=2$. Stable and
unstable solutions correspond to thick blue and thin red segments of the
lines, respectively. The dotted line in (c) represents the analytical
approximation given by Eq. (\protect\ref{stXc}). (d) The critical norm
(thick green line), at which the SSB\ takes place in the presence of bounded
pseudopotential (\protect\ref{eq:bounded}), vs. depth $A$ of its profile. At
$N<N_{\mathrm{cr}}$, no asymmetric modes exist, and symmetric ones are
stable. At $N>N_{\mathrm{cr}}$ there exist stable asymmetric modes, while
symmetric ones are unstable. The thin dashed and dotted lines depict
analytical predictions (\protect\ref{largeA}) and (\protect\ref{Ncr}), which
are valid for large $A$ and large $N_{\text{cr}}$, respectively. Indefinite
coefficient in Eq. (\protect\ref{largeA}) is chosen as $\mathrm{const}%
\approx 7$ to provide for the best fit. (e) and (f) Typical examples of
symmetric and asymmetric modes at $A=2$ and $\protect\mu =-2$. The thin
dotted lines depict the parabolic potential and pseudopotential $P(x)$. For
all panels, the strength of the HO trapping is $\protect\omega =\protect%
\sqrt{2}$.}
\label{fig:bif01}
\end{figure}

The resulting bifurcation diagrams for the bounded pseudopotential (\ref%
{eq:bounded}) are displayed in Fig.~\ref{fig:bif01}. As follows from Fig.~\ref{fig:bif01}(a,b),
a direct bifurcation of the pitchfork type \cite{bif} occurs, with the
increase of $N$, at the critical point, $N=N_{\mathrm{cr}}$, as shown in
panel (c): two asymmetric states, with $X_{c}>0$ and $X_{c}<0$, branch off
from the symmetric one, with $X_{c}=0$, at $N>N_{\mathrm{cr}}$. Because the
governing equation (\ref{1D_GP}) {keeps the global symmetry,} i.e.,
it is invariant with respect to the space reflection, $x\rightarrow -x$,
asymmetric solutions always emerge in two mutually-mirrored copies, i.e., $%
u(x)$ and $u(-x)$ with identical norms, therefore states $u(x)$ and $u(-x)$
are indistinguishable in the $(N,\mu )$-diagram plotted in Fig.~\ref%
{fig:bif01}(b). The SSB bifurcation generates asymmetric states above a
critical value of the norm, $N>N_{\mathrm{cr}}$. At $\mu \rightarrow 1$, the
norm $N$ is vanishing, and the symmetric mode transforms into the
harmonically trapped linear one with {an infinitely } small amplitude. Another
visualization of the SSB bifurcation is presented in Fig.~\ref{fig:bif01}%
(c), in the $(X_{c},N)$ plane.

Dependence of critical norm $N_{\mathrm{cr}}$ at the SSB point on depth $A$
of the modulation of pseudopotential (\ref{eq:bounded}) is the most
important characteristic of the setting under the consideration. The
numerically found dependence is plotted in Fig.~\ref{fig:bif01}(d), where
one observes that $N_{\mathrm{cr}}$ slowly decays at $A\rightarrow \infty $,
and $N_{\mathrm{cr}}$ diverges as $A$ decreases, in agreement with the
well-known fact that SSB does not occur in the model combining the linear HO
potential and spatially uniform self-attractive nonlinearity \cite%
{AZ07,ZAKP08,parabolic}. These asymptotic features can be easily explained
analytically. Indeed, in the case of very large $A$, term $1$ in definition (%
\ref{eq:bounded}) for $P(x)$ may be neglected, which allows one to remove $A$
by rescaling, leading to an asymptotic relation valid for $A\rightarrow
\infty $:
\begin{equation}
\left( N_{\mathrm{cr}}\right) _{A\rightarrow \infty }=\mathrm{const}\cdot
A^{-1}.  \label{largeA}
\end{equation}

Profiles of symmetric and asymmetric modes coexisting at the same value of
chemical potential $\mu $ are plotted in Fig.~\ref{fig:bif01}(e) and Fig.~%
\ref{fig:bif01}(f), respectively. Both solutions have the nodeless
single-peak shape, but the maximum of the symmetric mode is located exactly
at $x=0$, whereas the maximum of the asymmetric one is shifted to $x>0$, the
mirrored asymmetric solution having its maximum shifted to $x<0$.

Next, we address the linear stability of the found symmetric and asymmetric
solutions. First, Fig. \ref{fig:bif01}(b) clearly shows that both the
symmetric and asymmetric branches satisfy the necessary stability condition
in the form of the Vakhitov-Kolokolov (VK) criterion, $dN/d\mu <0$ \cite%
{VK,Berge,Fibich}. Then, following the standard procedure, the stability
problem amounts to the evaluation of eigenvalues of the linearization
operator $\mathcal{L}=\mathcal{L}_{+}\mathcal{L}_{-}$, where (see, e.g.,
book \cite{JYang} for details)
\begin{equation}
\mathcal{L}_{\pm }=\frac{d^{2}}{dx^{2}}+\mu -\frac{1}{2}\omega
^{2}x^{2}+(2\pm 1)P(x)u^{2}.
\end{equation}%
Mode $u(x)$ is linearly stable if all eigenvalues of $\mathcal{L}$ are real
and positive. If a negative or complex eigenvalue $\Lambda $ is found in the
spectrum of $\mathcal{L}$, then $u(x)$ is unstable, with instability growth
rate $|\text{Im }\sqrt{\Lambda }|$. Numerically computing the eigenvalues
for the symmetric and asymmetric states, we observe that the symmetric
branch is stable at $\mu >\mu _{\mathrm{cr}}$ (in other words, at $N<N_{%
\mathrm{cr}}$), where $\left( \mu _{\mathrm{cr}},N_{\mathrm{cr}}\right) $
are coordinates of the SSB bifurcation point in Fig. \ref{fig:bif01}. At $%
\mu <\mu _{\mathrm{cr}}$ ($N>N_{\mathrm{cr}}$), the symmetric mode is
destabilized by a single negative eigenvalue in the spectrum of $\mathcal{L}$%
, while the emerging asymmetric modes are \emph{stable}.

\begin{figure}[tbp]
\includegraphics[width=0.75\columnwidth]{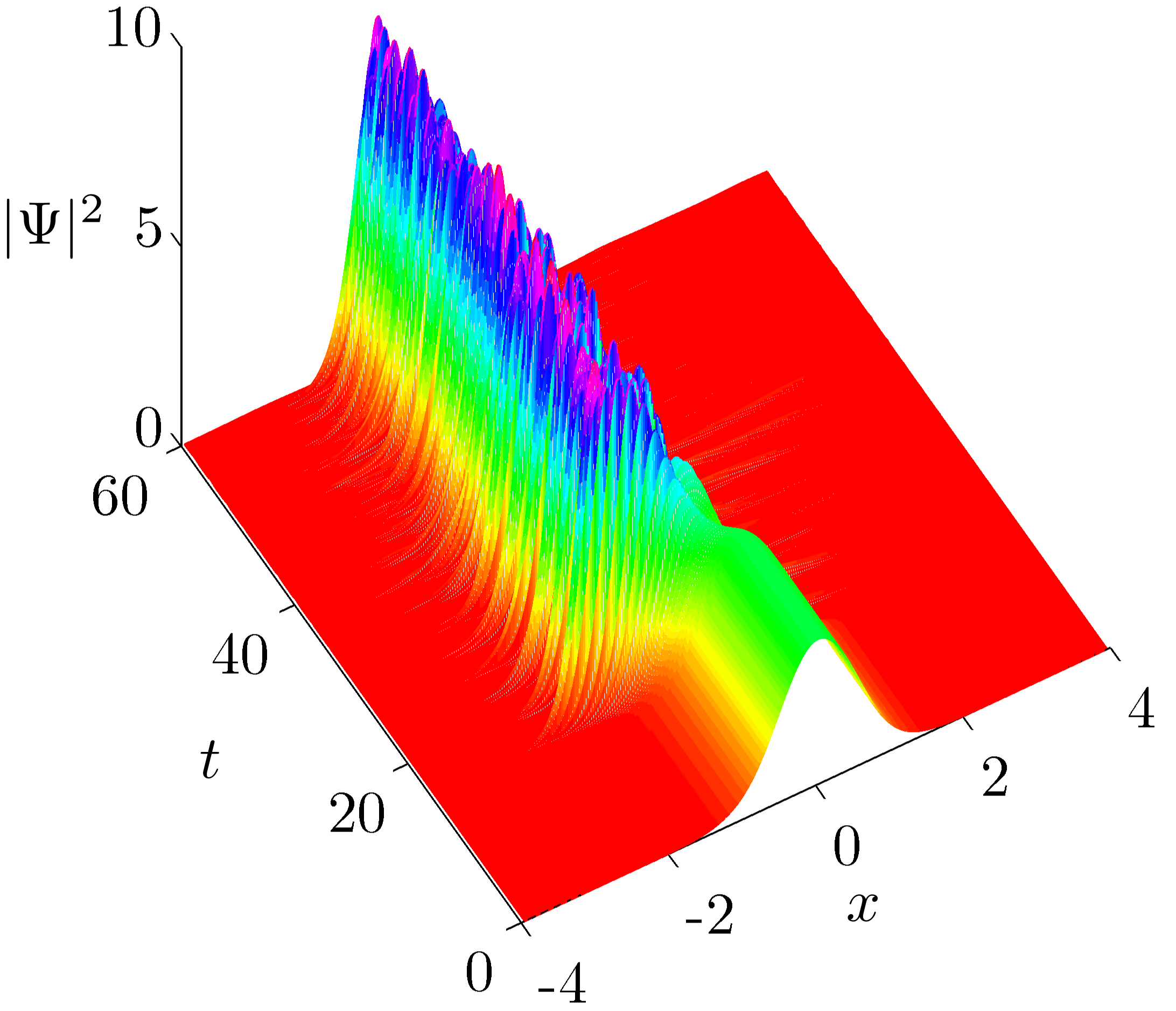}
\caption{Evolution $|\Psi (x,t)|^{2}$ of an unstable symmetric mode in
bounded pseudopotential (\protect\ref{eq:bounded}) with $A=2$, $\protect\mu %
=-2$ and $N\approx 4.90$.}
\label{fig:dyn01}
\end{figure}

Direct simulations of the evolution of symmetric and asymmetric modes in the
framework of time-dependent GPE (\ref{1D_GP}) confirm the predictions of the
linear-stability analysis. As a representative example, in Fig.~\ref%
{fig:dyn01} we display the behavior of an unstable symmetric mode. The input
was chosen in the form of the unstable mode, $u(x)$, with a small initial
perturbation added to it [in particular, an appropriate perturbation can be
introduced merely by multiplying $u(x)$ by $1.001$]. Even this mild (and
symmetric) perturbation rapidly triggers strong dynamical instability, which
spontaneously breaks the symmetry of the initial mode, shifting sideward (to
the left, in Fig. \ref{fig:dyn01}), thus creating an asymmetric state with
irregular internal oscillations, which do not destroy the emerging
asymmetric mode.

Now we aim to demonstrate that the SSB\ scenario can be essentially altered
by taking another pseudopotential, which is also expulsive, but with an
\emph{unbounded} shape, unlike the bounded one in Eq. (\ref{eq:bounded}):%
\begin{equation}
P(x)=1+Ax^{2}.  \label{eq:unbounded}
\end{equation}%
{Strictly speaking, the indefinite growth of the local nonlinearity
strength at }$|x|\rightarrow \infty ${, which is implied by Eq. (\ref%
{eq:unbounded}), is not possible, but in practical terms it may be truncated
at values of }$|x|${\ which are much larger than the size of the
trapped mode \cite{Borovkova}. In the physical settings, this
pseudopotential can be implemented by means of the same approach as the one
defined by Eq. (\ref{eq:bounded}).}

\begin{figure}[tbp]
\includegraphics[width=0.99\columnwidth]{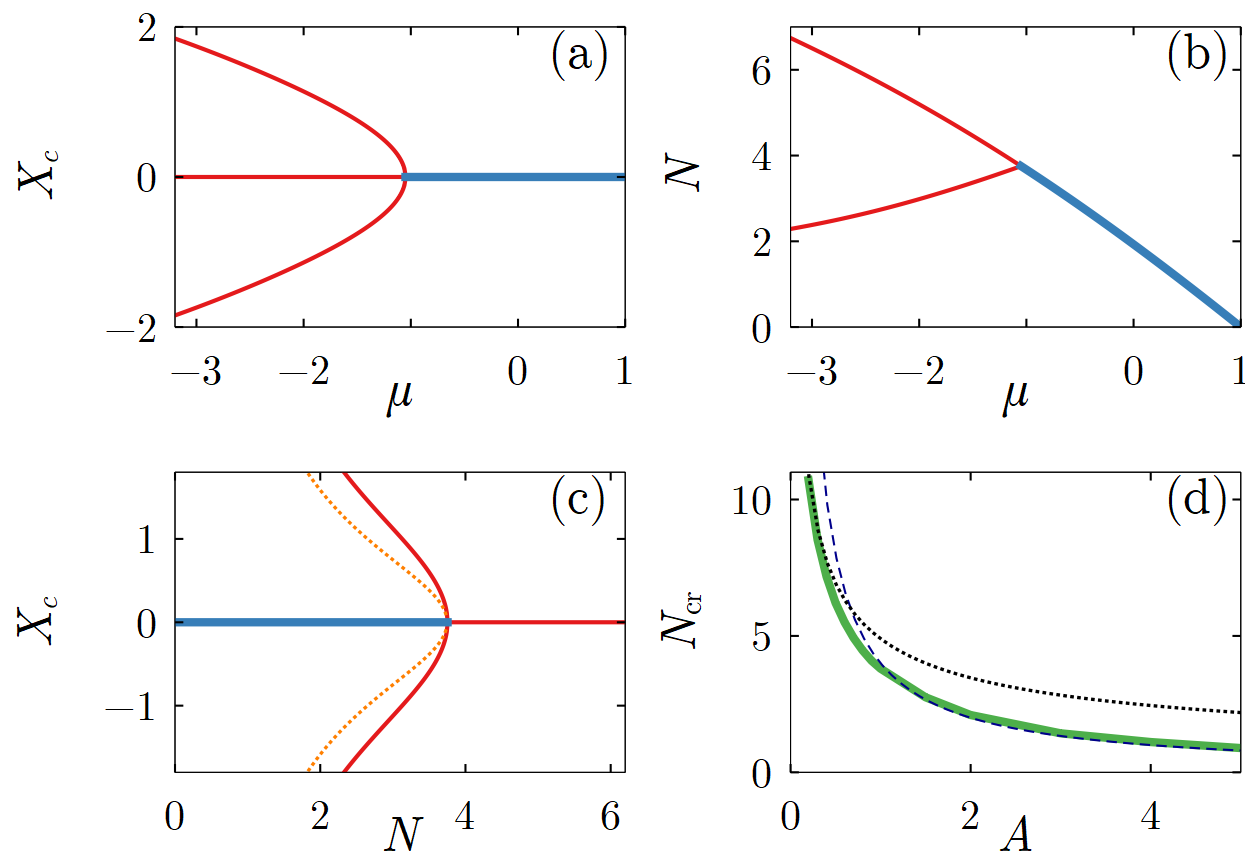}
\caption{The SSB bifurcation for the unbounded pseudopotential (\protect\ref%
{eq:unbounded}), with $A=1$ and $\protect\omega =\protect\sqrt{2}$. Panels
(a-d) are organized as in Fig.~\protect\ref{fig:bif01}. Dotted line in (c)
is the analytical prediction (\protect\ref{eq:Xunb}). For the analytical
approximation (\protect\ref{largeA}), valid for large $A$ [the dashed line
in (d)], $\mathrm{const}\approx 5$ is chosen to provide for the best fit. }
\label{fig:bif02}
\end{figure}

On the contrary to the case of pseudopotential (\ref{eq:bounded}), the
symmetric state undergoes the \textit{inverted} pitchfork SSB bifurcation
with the increase of $N$. The corresponding SSB diagrams are displayed in
Fig.~\ref{fig:bif02}(a,b,c), where the panels with different renditions of
the SSB bifurcation are organized in the same way as in Fig.~\ref{fig:bif01}%
. In particular, we observe that slope $dN/d\mu $ of the asymmetric branch
is positive in this case, hence the VK criterion implies that the branch is
unstable. This prediction agrees with the numerical computation of the
linear stability eigenvalues, which produces a single negative (unstable)
eigenvalue in the spectrum of the linearization operator $\mathcal{L}$ for
this branch. As above, the symmetric state is destabilized by the
bifurcation (again, with one negative eigenvalue emerging in the spectrum of
$\mathcal{L}$), when the pair of the unstable asymmetric states merge into
the symmetric one, at $N=N_{\mathrm{cr}}$. Furthermore, at $N<N_{\mathrm{cr}}
$ the symmetric state is not a ground state, but only a metastable one,
which is made clear by both numerical simulations and analytical results
presented below. The observed change of the slope of $N(\mu )$ dependence
and the destabilization of the symmetric and asymmetric modes is consistent
with the rigorous treatment developed in Ref. \cite{Yang13}. Spatial shapes
of solutions in the unbounded potential are qualitatively similar to those
in the bounded one [see Fig.~\ref{fig:bif01}(e,f)], therefore they are not
shown in Fig.~\ref{fig:bif02}.

\begin{figure}[tbp]
\includegraphics[width=0.75\columnwidth]{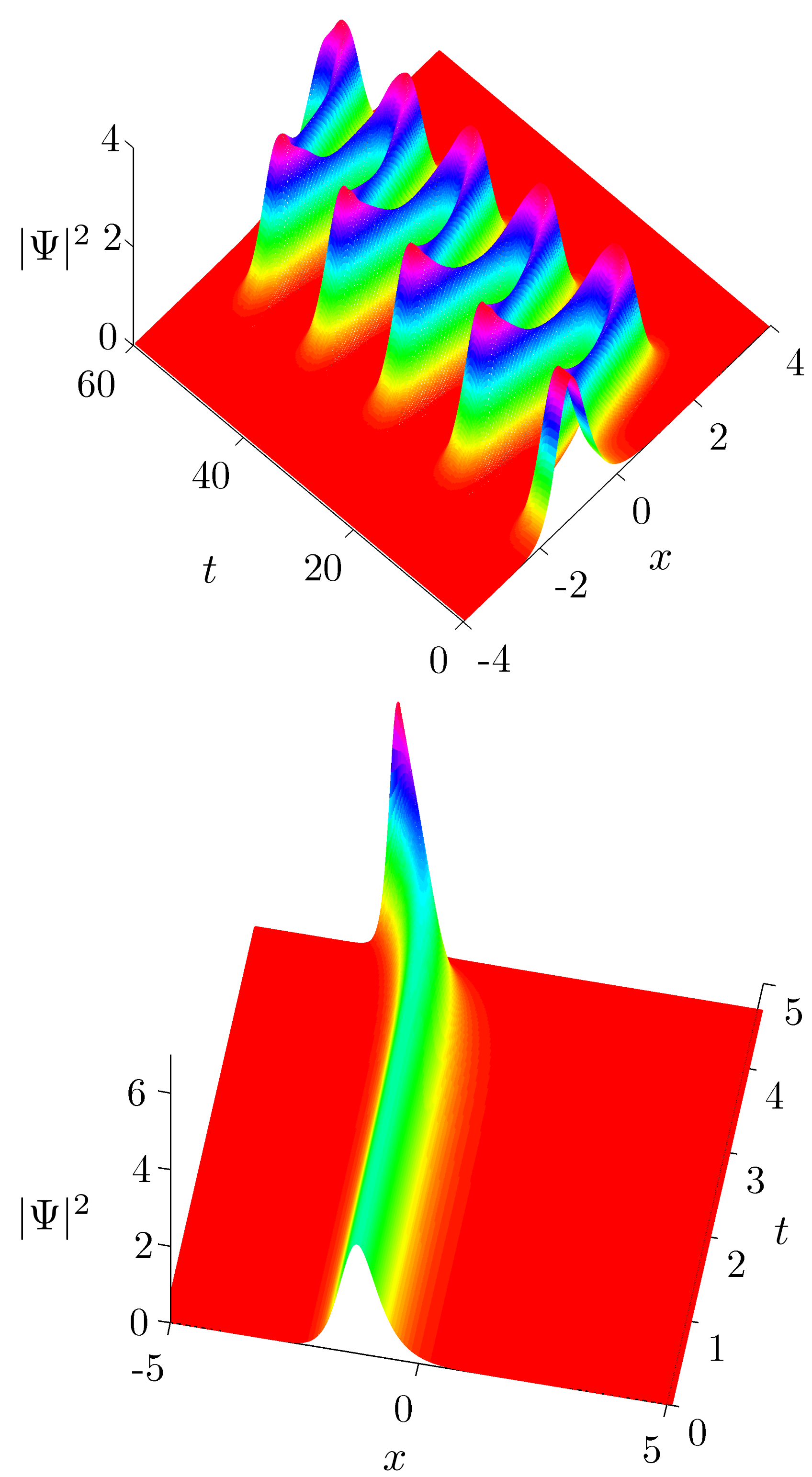}
\caption{Different dynamics for the same unstable asymmetric mode, with $%
\protect\mu =-2$, subjected to different initial perturbations, under the
action of unbounded pseudopotential (\protect\ref{eq:unbounded}) with $A=1$.}
\label{fig:dynx2}
\end{figure}

Dynamics of unstable asymmetric states in the unbounded pseudopotential is
sensitive to the choice of initial perturbations. In particular, multiplying
an unstable asymmetric mode $u(x)$ with $X_{c}<0$ by $e^{ikx}$, with
right-directed \textit{kick} $0<k\ll 1$, one triggers oscillations of the
solution's center between $X_{c}$ and $-X_{c}$, as shown in Fig.~\ref%
{fig:dynx2}(a). On the other hand, the application of small $k<0$ naturally
initiates drift of the quasi-soliton further to the left, where the growing
attractive nonlinearity makes the solution very narrow, driving the growth
of its amplitude, see Fig.~\ref{fig:dynx2}(b). A similar scenario, i.e.,
spontaneous sideward drift, is typical for unstable symmetric modes.

\section{The analytical approach}
\label{sec:analytical}
In addition to the simple analytical
result given by Eq. (\ref{largeA}), which is relevant for small $N$, an
analytical approximation can be developed for large $N$. In this case,
strong nonlinearity makes the soliton, with its center located at a point
with coordinate $X$, very narrow, hence Eq. (\ref{eq:stat}) gives rise to
the approximate solution,
\begin{equation}
u_{0}\left( x\right) =\sqrt{-\frac{2\mu }{P(X)}}\mathrm{sech}\left( \sqrt{%
-\mu }\left( x-X\right) \right) ,~N=\frac{4\sqrt{-\mu }}{P(X)},
\label{u0(xi)}
\end{equation}%
\ Then, the substitution of this in Eq. (\ref{H}) yields an effective
soliton's potential energy,
\begin{equation}
U(X)=-(N^{3}/48)\left( P(X)\right) ^{2}+\left( \omega ^{2}N/2\right) X^{2},
\label{U}
\end{equation}%
which predicts equilibria, $X=X_{c}$ [see Eq. (\ref{Xc})], at points with $%
U^{\prime }(X_{c})=0$.

First, for vanishingly small $X$, the expansion of potential (\ref{U})
yields, for both forms of $P(A)$ defined by Eqs. (\ref{eq:bounded}) and (\ref%
{eq:unbounded}),
\begin{equation}
U(X)=-(N^{3}/48)+(N/2)\left[ \omega ^{2}-(AN^{2}/12)\right] X^{2}+\mathcal{O}%
(X^{4}).
\end{equation}%
From here it follows that the SSB bifurcations, which implies a transition
from the stable to unstable equilibrium at $X=0$, occurs at%
\begin{equation}
N_{\mathrm{cr}}=2\omega \sqrt{3/A}.  \label{Ncr}
\end{equation}

Further, at $N>N_{\mathrm{cr}}$ the substitution of the bounded
nonlinearity-modulation profile (\ref{eq:bounded}) in equation $%
dU/dX|_{X=X_{c}}=0$, which follows from Eq. (\ref{U}), {yields an
equation predicting a mutually symmetric pair of the asymmetric equilibrium
points,}%
\begin{equation}
\left( 1+A\tanh ^{2}X_{c}\right) \frac{\tanh X_{c}}{\cosh ^{2}X_{c}}=\frac{%
12\omega ^{2}}{AN^{2}}X_{c}.  \label{stXc}
\end{equation}%
Then, straightforward analysis demonstrates that the asymmetric equilibria
are always\textit{\ local minima} of potential (\ref{U}), with $%
d^{2}U/dX^{2}|_{X=X_{c}}>0$, hence {this pair of the equilibrium
points} are stable, in accordance with the numerical findings presented
above.

The validity of approximations (\ref{Ncr}) and (\ref{stXc}) for the bounded
pseudopotential is illustrated in Figs.~\ref{fig:bif01}(d) and \ref%
{fig:bif01}(c), respectively. For sufficiently large $N$, the analytical
predictions (dotted lines) are practically identical to their numerical
counterparts.

For the unbounded modulation profile (\ref{eq:unbounded}), the equilibrium
condition predicts the existence of unstable asymmetric equilibria [local
potential \textit{maxima}, with $d^{2}U/dX^{2}|_{X=X_{c}}<0$] in the
subcritical region, at $N<N_{\mathrm{cr}}$:
\begin{equation}
X_{c}^{2}=A^{-1}\left( N_{\mathrm{cr}}^{2}/N^{2}-1\right) ,  \label{eq:Xunb}
\end{equation}%
which explains the numerical findings reported above, see the comparison
with numerical results in Fig. \ref{fig:bif02}(c,d). Furthermore, effective
potential (\ref{U}) is always \textit{globally expulsive} in the present
case, with term $-\left( N^{3}/48\right) A^{2}X^{4}$ dominating at large $|X|
$. This fact implies that the model with the unbounded nonlinear potential
does not have a ground state, the symmetric one being \textit{metastable} at
$N<N_{\mathrm{cr}}$, as mentioned above.

\section{Conclusion}
\label{sec:conclusion}
For the first time, to the best of our knowledge, we
have demonstrated that the competition between \emph{\ single-well} linear
and nonlinear potentials enables the effect of the SSB (spontaneous symmetry
breaking) of nonlinear modes. The reported SSB\ bifurcation scenario is
rather unusual since, contrary to most of the previously reported settings
where the symmetry breaking has been encountered, our system does not
require any double- or multi-well potential. With the increase of the
solution's norm, $N$, the bifurcation occurs at the critical value, $N=N_{%
\mathrm{cr}}$. There two different bifurcation scenarios, depending on the
form of the nonlinear pseudopotential. If it is bounded, the bifurcation is
direct, destabilizing the symmetric ground state and producing the pair of
asymmetric ones. On the other hand, the bifurcation is inverted, leading to
the merger of the pair of unstable asymmetric states into the symmetric one,
for unbounded pseudopotentials. Parallel to the systematically collected
numerical results, basic results are accurately explained by the simple
analytical approximation, which represents the self-trapped modes as narrow
solitons. The settings explored in this work may be realized in BEC and
nonlinear optics.

As an extension of the present analysis, it may be relevant to consider a
setting with the competition between a confining nonlinear potential and
expulsive linear one, {as a generalization of the antiwaveguiding
settings in the self-focusing media}; in particular, the analytical
approximation developed here may be relevant in that case too. A challenging
possibility is to consider a two-dimensional version of the present model,
which, as well as the one-dimensional system, can be realized in BEC and
nonlinear optics (using bulk waveguides, in the latter case). Finally, while
the present work is focused on the symmetry breaking of the nodeless ground
states, it is also relevant to extend the analysis for the excited states,
in which SSB effects may be expected too.

\begin{acknowledgments}
Authors are grateful to K.~A.~Kolesnikova for
the help in numerical computations. The research of G.L.A, M.E.L and D.A.Z.
was supported by  Russian Science Foundation (Grant No. 17-11-01004).
The work of D.A.Z. was also supported by Government of Russian Federation (Grant 08-08).
The work of B.A.M. is supported, in part, by Israel Science Foundation (Grant No.
1287/17).
\end{acknowledgments}

\end{document}